\UseRawInputEncoding

\documentclass[letterpaper, 10 pt, conference]{ieeeconf}  % Comment this line out if you need a4paper

\IEEEoverridecommandlockouts                              % This command is only needed if 
\usepackage{graphics} % for pdf, bitmapped graphics files
\usepackage{epsfig} % for postscript graphics files
\usepackage{mathptmx} % assumes new font selection scheme installed
\usepackage{times} % assumes new font selection scheme installed
\usepackage{amsmath} % assumes amsmath package installed
\usepackage{amssymb}  % assumes amsmath package installed
\usepackage{multirow}
\usepackage{subfigure}
\usepackage{cite}

\title{\LARGE \bf
ExploreGS: a vision-based low overhead framework for 3D scene reconstruction
}

\author{Yunji Feng$^{}$, Chengpu Yu$^{*}$, Fengrui Ran$^{}$, Zhi Yang$^{}$, Yinni Liu$^{}$% <-this % stops a space\
\thanks{This work was supported by the Chongqing Natural Science Foundation CSTB2023NSCQ-JQX0018,  National Natural Science Foundation of China (Grant No. 62473046, No. 62088101), and Beijing Natural Science Foundation L221005.}% <-this % stops a space
\thanks{All authors are with School of Automation, Beijing Institute of Technology, Beijing, China and Chongqing innovation Center, Beijing Institute of Technology, Chongqing, China {\tt\small \{fengyunji@bit.edu.cn, yuchengpu@bit.edu.cn, ranfengrui@bit.edu.cn, 3220241334@bit.edu.cn, 1120212479@bit.edu.cn\}}
}
}

\begin{document}

\maketitle
\thispagestyle{empty}
\pagestyle{empty}

%%%%%%%%%%%%%%%%%%%%%%%%%%%%%%%%%%%%%%%%%%%%%%%%%%%%%%%%%%%%%%%%%%%%%%%%%%%%%%%%
\begin{abstract}

%This paper proposes a low-overhead, vision-based 3D reconstruction framework for drones called ExploreGS. By using only RGB images, ExploreGS replaces traditional lidar-based point cloud acquisition with a visual model, reducing costs while maintaining high-quality reconstruction. The framework integrates scene exploration and viewpoint planning, employs a Bag-of-Words model for real-time inference, and completes 3D Gaussian Splatting (3DGS) training onboard. Experiments in simulation and real-world scenarios demonstrate its efficiency and applicability on resource-constrained devices, matching state-of-the-art methods in reconstruction quality. ExploreGS offers a cost-effective and flexible solution for 3D scene reconstruction in unknown environments.

This paper proposes a low-overhead, vision-based 3D scene reconstruction framework for drones, named ExploreGS. By using RGB images, ExploreGS replaces traditional lidar-based point cloud acquisition process with a vision model, achieving a high-quality reconstruction at a lower cost. The framework integrates scene exploration and model reconstruction, and leverags a Bag-of-Words(BoW) model to enable real-time processing capabilities, therefore, the 3D Gaussian Splatting (3DGS) training can be executed on-board. Comprehensive experiments in both simulation and real-world environments demonstrate the efficiency and applicability of the ExploreGS framework on resource-constrained devices, while maintaining reconstruction quality comparable to state-of-the-art methods.

\end{abstract}

%%%%%%%%%%%%%%%%%%%%%%%%%%%%%%%%%%%%%%%%%%%%%%%%%%%%%%%%%%%%%%%%%%%%%%%%%%%%%%%%
\section{INTRODUCTION}

3D scene reconstruction has found extensive applications in various fields, including urban planning, digital asset acquisition, and structural inspection. In recent years, drone technology has emerged as a critical tool for large-scale 3D reconstruction data acquisition, due to its inherent agility and versatility in diverse environmental conditions. These aerial systems are capable of efficiently covering comprehensive scene areas, overcoming the limitations posed by challenging terrains and complex site conditions, thereby significantly advancing spatial data collection workflows. % With the advancement of computer graphics, traditional reconstruction methods such as point cloud and voxel-based approaches have limitations in effectively representing object details and scale. Deep learning-based methods have gradually become the dominant approach in 3D reconstruction. Among these, 3D Gaussian Splatting (3DGS) \cite{3DGS} have gained significant attention in robotics due to its explicit representation capability.

In recent studies, numerous data acquisition and 3D scene reconstruction methods have been proposed. However, the majority of these reconstruction methods involve complex data acquisition processes, impose high demands on computational equipment performance, and face challenges in practical experimental validation. First, many methods require drones to perform pre-modeling of the target scene beforehand, obtaining a rough 3D model, and then planning flight paths based on this model for secondary flights to collect detailed data. Second, due to the significant fluctuations of the collected depth data in real-world scenarios, most methods cannot rely solely on camera for scene reconstruction, necessitating additional point cloud information acquisition through lidar, thereby increasing task costs. Additionally, reconstruction methods that depend on visual data typically employ Structure from Motion (SfM) to obtain prior point cloud models and camera extrinsics. However, this approach experiences exponential growth in computational overhead as the number of images increases, making it difficult to deploy on edge computing devices. On the other hand, these methods are unable to estimate camera extrinsics in scenes with similar textures or insufficient image overlap shown in Fig.1.

\begin{figure}[htbp]
        \centering
        \includegraphics[width=0.9\linewidth]{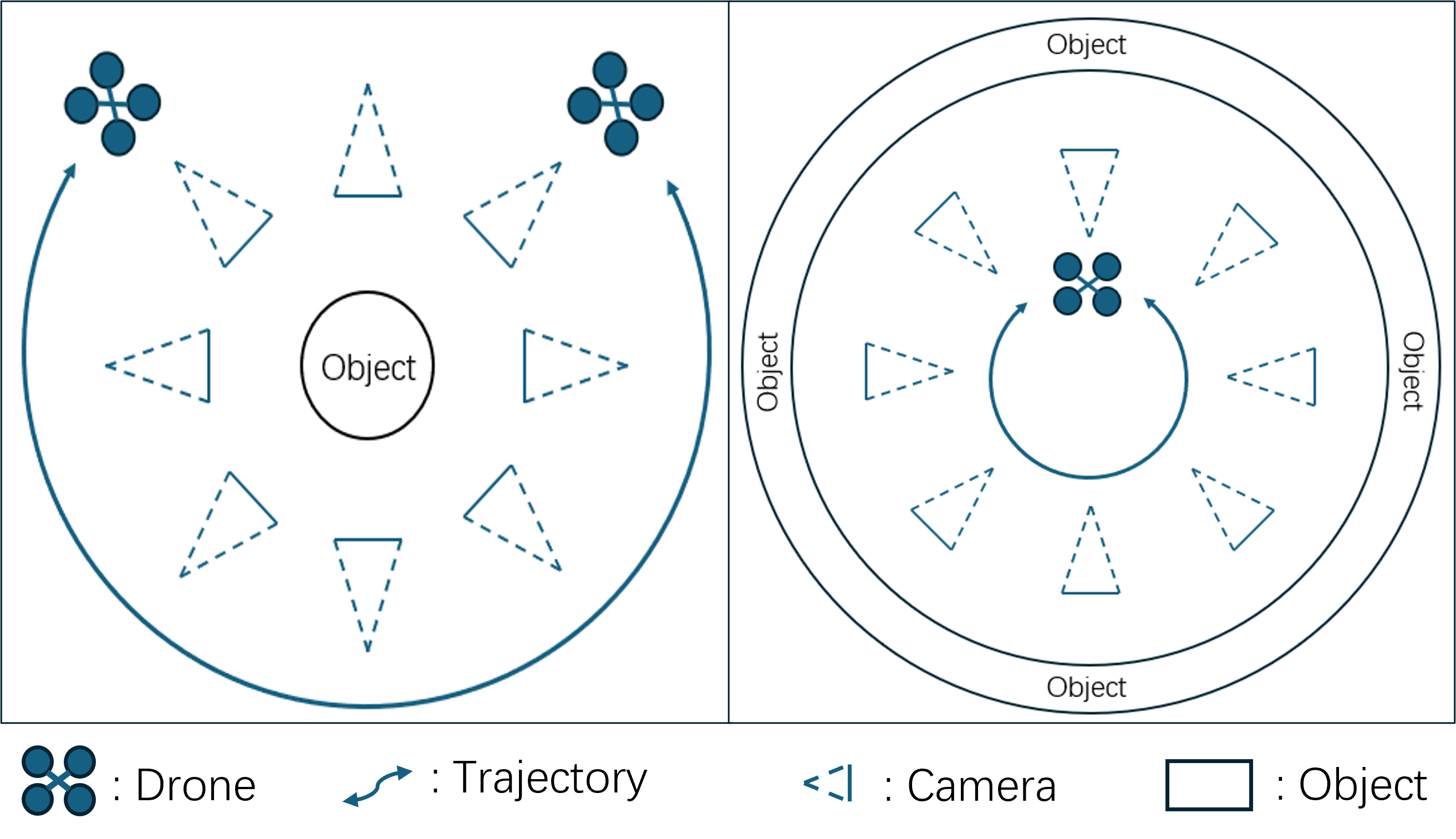}
        \caption{Traditional SfM algorithms can achieve good results in scenes with a single object (left), but they usually fail in scenes where the overlap between images in the sequence is small (right)}
\end{figure}

In this paper, we present an autonomous drone-based 3D reconstruction framework that exclusively leverages camera for 3DGS-based scene reconstruction on-board. Our approach integrates a visual model to supplant traditional lidar-based methods for point cloud generation, while incorporating an enhanced autonomous exploration strategy to minimize the required image data. It enables real-time reconstruction on edge computing device, culminating in the training and rendering of 3DGS models. First, the drone autonomously explores an unknown environment, using a depth camera to construct a rough 3D model only for real-time path planning and obstacle avoidance, while simultaneously capturing RGB image data to perform a thorough scan of the scene's object surfaces. Meanwhile, a fast and efficient image pairing selector identifies a minimal subset of images as keyframes from the sequence and pairs images that have overlapping fields of view, resulting in a series of image pairs. These image pairs are then fed into a visual model to generate local point cloud models, which are subsequently aligned through global registration to obtain a complete scene point cloud model. Finally, with only a few training iterations, a high-quality 3DGS model can be achieved.

% We compared our method with classical approaches. The results show that our method achieves better reconstruction quality in much shorter time (5 times faster). In summary, the contributions of this paper are as follows:
We compare the proposed method with classic approaches through both numerical simulations and real-world experiments. The results show that it achieves fairly good reconstruction quality in significantly less time (5 times faster than existing methods). In summary, the key contributions of this paper are as follows:

\begin{itemize}

\item  A fast vision model based data acquisition method is proposed which efficiently explores unknown environments and generates point cloud.

\item A pair selection scheme is designed based on the BoW model, which can reduce redundant data by $84-95\%$.

\item  The fully autonomous reconstruction framework is validated in both simulation and real-world scenarios, enabling real-time inference on-board. 

\end{itemize}

The structure of this paper is organized as: Section II reviews the related work, covering autonomous data acquisition and 3D reconstruction methods. Section III details the proposed ExploreGS framework for 3D scene reconstruction. Section IV provides the experimental setting and results, including both simulation and real-world tests. Finally, Section V concludes the paper and outlines future research directions.

\section{RELATED WORK}

\subsection{Autonomous data acquisition}

Drone-based autonomous data acquisition uses drone-mounted sensors, such as lidar and cameras, to perceive environmental information within a scene. The system plans the shortest coverage path and then controls the drone to follow the path to acquire point cloud and image data. Based on different task requirements, this technology can be categorized into two types of planning algorithm: model-based algorithm and model-free algorithm. 

The model-based method typically relies on pre-existing point cloud models or grid maps of the scene. On this basis, the system performs offline planning to determine optimal viewpoints for drone-based image acquisition. \cite{Plan3D,Real-Time-UAV} evaluates the completeness of the existing model and replans the path to address missing areas. %\cite{Plan3D,Real-Time-UAV} plan path within designated areas, using SLAM algorithms to establish an initial model from an overhead view, and then performs secondary data collection. \cite{Offsite-aerial} plans paths based on satellite maps, while \cite{Sampling-Based-Path} manually plans paths to collect data. 
\cite{FCPlanner} performs spatial decomposition based on a skeleton framework for existing 3D models and completes path planning based on this skeleton structure.

The model-free method, also known as the real-time decision-making planning method, dynamically generates optimal trajectories through continuous environmental perception. \cite{Next-Best-View} proposes an Next-Best-View (NBV) selection framework that maximizes the entropy of information to compute collision-free coverage paths for incremental area modeling. \cite{Active3D} develops an online feedback loop integrating SLAM with Multi-View Stereo(MVS) depth estimation for simultaneous mapping and path planning. %\cite{Next-Best-View} employing TSDF volumetric representation with clustering-based view path optimization for multi-UAV collaborative exploration; \cite{ViewPathPlanning} implementing surface completeness assessment through quality-aware gap detection to formulate inspection trajectories balancing MVS performance and coverage completeness. 
\cite{PredRecon} establishes surface prediction models to guide exploration of unobserved regions, collectively advancing autonomous reconstruction through adaptive perception-planning integration.

\subsection{3D Reconstruction}

In robotic perception systems, the RGB-D camera offers cost-effective alternatives to LiDAR with compact form factors and reduced payload requirements, particularly for point cloud acquisition as demonstrated in \cite{GS-Planner}. But practical deployments reveal significant depth-noise challenges that limit complete sensor substitution. Current monocular 3D reconstruction methodologies using RGB data can be categoried into three distinct paradigms: Structure-from-Motion (SfM), MVS and learning based methods. SfM leverages feature point extraction \cite{KeypointDetection} and pairwise image registration \cite{HSfM,Structure-from-Motion-Revisited} for sparse reconstruction through epipolar geometry optimization. MVS systems \cite{AdaptivePathDeformation} performs dense matching via camera parameters derived from SfM to establish photometric consistency constraints. Most learning-based approaches employ pixel-to-point regression architectures, including monocular depth estimation frameworks \cite{VisionTransFormersForDense,DepthAnything} and multi-view geometric reconstruction models \cite{DUSt3R,MASt3R}.

% Recent advancements in 3D reconstruction have witnessed paradigm-shifting scene representations exemplified by Neural Radiance Fields (NeRF) and 3D Gaussian Splatting (3DGS). NeRF \cite{NeRF} establishes differentiable implicit scene modeling through coordinate-based MLPs that parameterize volumetric radiance fields, achieving photorealistic novel view synthesis vias neural rendering. In contrast, 3DGS \cite{3DGS} employs explicit geometric primitives through anisotropic 3D Gaussians, enabling real-time differentiable rendering via splatting-based rasterization. This explicit representation demonstrates superior adaptability for dynamic scene reconstruction, geometric manipulation, and physics-based simulation applications compared to NeRF's implicit paradigm. Our framework adopts 3DGS as the core reconstruction methodology, with rendered image quality quantified through Peak Signal-to-Noise Ratio (PSNR) metrics.
Recent advancements in 3D reconstruction have witnessed paradigm-shifting scene representations exemplified by 3D Gaussian Splatting (3DGS) \cite{3DGS}. 3DGS employs explicit geometric primitives through anisotropic 3D Gaussians, enabling real-time differentiable rendering via splatting-based rasterization. This explicit representation demonstrates superior adaptability for geometric manipulation and physics-based simulation. Our framework adopts 3DGS as the core reconstruction methodology, with rendered image quality quantified through Peak Signal-to-Noise Ratio (PSNR) metrics.

\begin{figure*}[t!]
	\centering
	\includegraphics[width=1.0\textwidth]{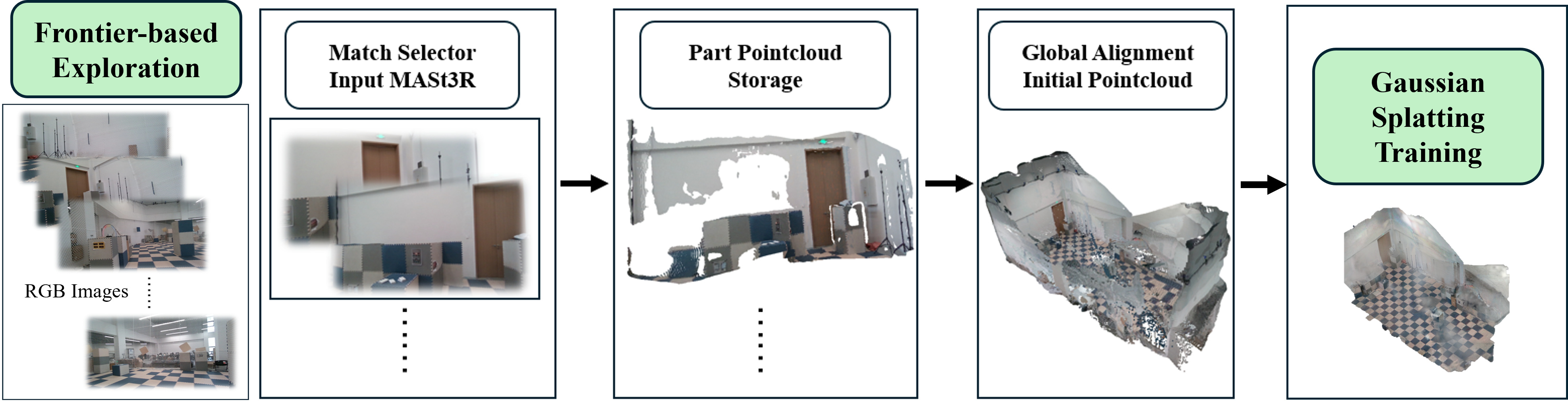}
	\caption{Overview of our ExploreGS framework. We collect RGB data while exploring the unknown scene. Given two images with common viewing area as input, we get part point cloud continuously and combine them after all images are processed. With an initial point cloud generated, few GS trainings can achieve good result.}
\end{figure*}

\section{Our Approach}

% We propose a novel 3D Gaussian reconstruction framework named ExploreGS, which is capable of real-time computation on the drone side. The overall framework of the algorithm is shown in the Fig.2. The goal of the algorithm is to enable a drone equipped with only a Realsense camera to autonomously capture photos in an unknown scene without external guidance and to complete the training of the Gaussian Splatting model on a platform with limited resources.

We propose a novel 3D reconstruction pipeline named ExploreGS. The overall framework is illustrated in Fig.2. The workflow commences with the drone's autonomous exploration of the unknown environment, during which it captures sequential RGB images. Meanwhile, the Match Selector identifies image pairs that have a common view area, and a vision model will process these image pairs to generate corresponding point cloud. Finally, we can get a high-fidelity 3DGS model by only a minimal number of trainings on the initial point cloud.

\subsection{Data Acquisition}

In unknown environments, our system enables autonomous drone operation through integrated environment perception, mapping, and exploration planning. The framework processes depth-sensor data to construct an occupancy grid , where unexplored boundary regions (frontiers) are identified through neighbor-voxel analysis. As demonstrated in Fig.3, this approach inherently supports multi-frontier detection without requiring predefined environmental priors.

\begin{figure}[htbp]
        \centering
        \includegraphics[width=0.8\linewidth]{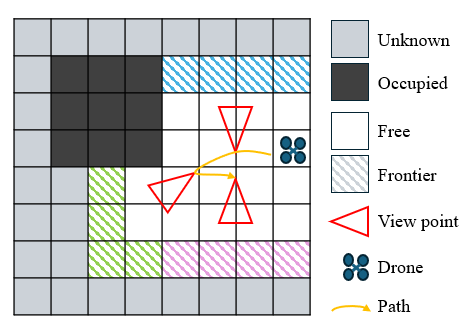}
        \caption{The quadrotor of real-world test. }
\end{figure}

We implement a hierarchical viewpoint planning framework combining dimensionality reduction and combinatorial optimization. First, Principal Component Analysis (PCA) projects frontier clusters into lower-dimensional manifolds to generate candidate viewpoints. Each viewpoint's utility is quantified through its coverage of the unknown area and the number of boundary points. Then, the Asymmetric Traveling Salesman Problem (ATSP) algorithm is used to sort the viewpoints \cite{FUEL}. For viewpoints $v_1$ and $v_2$, the corresponding directions are $yaw_1$ and $yaw_2$, the matrix for ATSP $M_{tsp}$ can be defined as:

$$
M_{tsp}(v_1,v_2)=\max \left\{ \frac{dis(v_1,v_2)}{v_{max}}, \frac{|yaw_1-yaw_2|}{yaw_{max}} \right\}
\eqno{(1)}
$$
where function $dis$ calculate the length of the collision-free path from $v_1$ to $v_2$. $v_{max}$ and $yaw_{max}$ are the linear and angular velocities of the drone.

Our trajectory generation module implements the framework of \cite{FastPlanner} to produce a smooth and dynamically feasible B-spline trajectory. By limiting $yaw_{max}$, we ensure stable visual data acquisition during the autonomous flight by maintaining the angular momentum within hardware limits. This dual optimization of spatial continuity and attitude stability enables persistent environment observation throughout the planned trajectory.

\subsection{Match Selector}

The Match Selector module efficiently identifies view-overlapping image pairs through a matching strategy inspired by \cite{VINS}. By implementing a BoW model \cite{DBoW} for rapid feature-similarity assessment, the system performs image retrieval from sequential data. % As shown in Fig.4, this approach balances computational efficiency with matching accuracy, enabling real-time operation while maintaining robust correspondence detection across overlapping viewpoints.

The framework processes $n$-frame image sequences ${Img_1, ..., Img_n}$ through three integrated stages: FAST corner detection \cite{FAST} extracts keypoints using Bresenham circle patterns on grayscale images, followed by BRIEF descriptors \cite{BRIEF} encoding geometric features into binary representations. A dynamically updated BoW model \cite{VINS} incrementally indexes these descriptors, enabling efficient feature matching through real-time database expansion. % This architecture achieves accelerated processing by leveraging FAST's pixel-intensity comparisons and BRIEF's bitwise operations, ensuring low time complexity for sequential image analysis.

For the pixel $p$ in the $k$-th image $Img_k$ with intensity $I_p$, the FAST detector \cite{FAST} identifies feature points. If the brightness of 12 consecutive points on the Bresenham circle is simultaneously greater than $I_p+t$ or less than $I_p-t$, then $p$ is considered a feature point, where $t$ is the threshold. % This deterministic intensity comparison leverages the Bresenham algorithm's discrete circular sampling pattern, enabling $O(1)$ computational complexity per pixel.

To describe image features concisely and accurately, the BRIEF descriptor utilizes binary encoding to efficiently characterize image features. Centered at the feature pixel $p$, an $S \times S$ neighborhood window is sampled, within which randomly selected point pairs $(x,y)$ undergo intensity comparisons. The binary test result $I_{p;x,y}$ is defined as:

$$
I_{p;x,y}=
\begin{cases}
        1, & I_x < I_y \\
        0, & I_x \geq I_y
\end{cases}
\eqno{(2)}
$$

By aggregating N such comparisons, the descriptor constructs a compact N-bit binary code $B \in \mathbb{R}^N$. % enabling efficient storage and matching while preserving discriminative power through localized intensity relationships.

The system utilizes hierarchical BoW for an efficient image matching. A pre-trained visual dictionary from VINS \cite{VINS} partitions the BRIEF descriptor space into $W$ visual words, forming a compact vocabulary. For the image $Img_k$, its descriptor $B_k$ is mapped through the dictionary's tree structure by iteratively selecting nodes that minimize the Hamming distance at each hierarchy level, ultimately generating the BoW vector $v_t \in \mathbb{R}^W$. % The similarity $s(v_1,v_2)$ between vectors $v_1$ and $v_2$ is defined as:

During processing the $k$-th image $Img_k$, candidate matching pairs ${<v_1,v_k>,...,<v_{k-1},v_k>}$ are evaluated. The similarity scores exhibit scene-dependent variations influenced by both image content and visual word distribution in the vocabulary. To ensure cross-scene consistency, we apply the dynamic normalization for the final score $\eta$:

$$
\eta (v_i,v_k)=\frac{s(v_i,v_k)}{s(v_{k-\Delta t},v_k)}
\eqno{(3)}
$$

$$
s(v_i,v_k)=1-\frac{1}{2}\left|{\frac{v_i}{|v_i|} - \frac{v_k}{|v_k|}}\right|
\eqno{(4)}
$$
where $s(v_i,v_k)$ is the similarity between $v_i$ and $v_k$. The historical BoW vector $v_{k-\Delta t}$ is retrieved through a dual-criterion selection mechanism: temporal proximity (constrained by  $\Delta t<T_{max}$ to ensure temporal continuity) and similarity filtering ($s(v_{k-\Delta t},v_k)>\tau$ where $\tau$ is the similarity floor threshold). %, effectively mitigating anomalous scoring caused by abrupt scene variations while preserving sequential matching stability.

To mitigate temporal redundancy in image sequences, we establish a similarity threshold $thr_{in}$ as the primary admission criterion for the database. The $k$-th image $Img_k$ is selected or discarded based on its similarity score with the last selected image $Img_{last}$, defined as:
$$
Img_k:
\begin{cases}
        Select, & \eta(v_{k-1},v_{last})<thr_{in} \\
        Discard, & \eta(v_{k-1},v_{last})\geq thr_{in}
\end{cases}
\eqno{(5)}
$$

As shown in Fig.4, the Match Selector generates pairwise vector correspondences $<v_i,v_k>,...,<v_j,v_k>$ whose scores $\eta$ are bigger than $\tau$, and their associated image matches are denoted by $<Img_i,Img_k>,...,<Img_j,Img_k>$. 

\begin{figure}[htbp]
        \centering
        \includegraphics[width=0.9\linewidth]{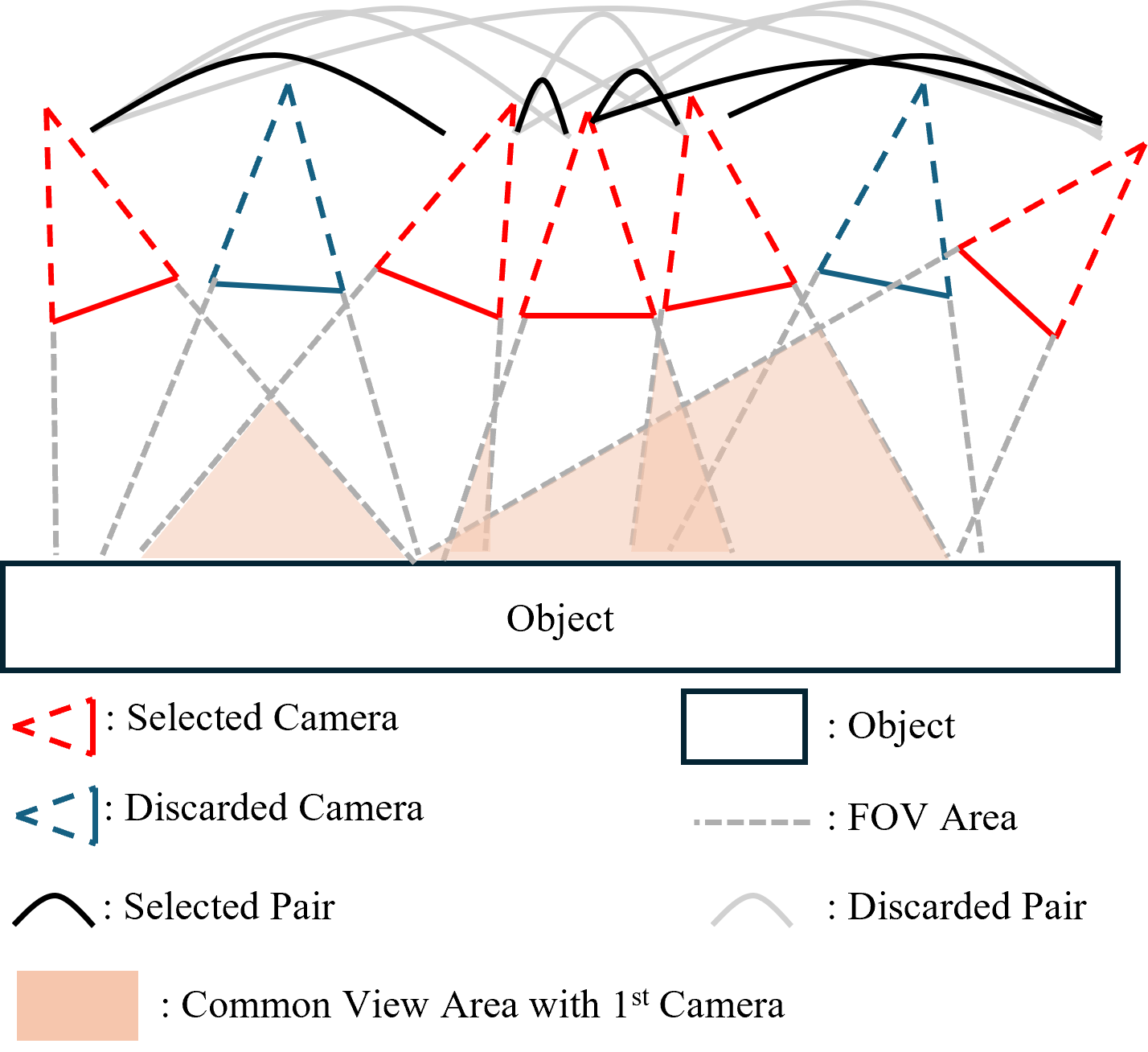}
        \caption{Selection method of image acquisition in flight. The number of image pairs can be greatly reduced by comparing the similarity of common view area and screening the pictures with too large similarity. }
\end{figure}

Through this optimized matching framework, the visual inference workload is reduced to critical feature pairs only, achieving significant computational efficiency improvements over baseline methods as quantified in Table 1.

\begin{table}[h]
\caption{The number of image pairs with different number of pictures}
\label{table_example}
\begin{center}
\begin{tabular}{|c||c|c|c|}
\hline
\multirow{2}*{Method} & \multicolumn{3}{|c|}{Number of Pictures} \\
\cline{2-4}
~ & 20 & 40 & 60 \\
\hline
\hline
Complete \cite{InstantSplat} & 380 & 1560 & 3540 \\
\hline
Swin & 100 & 200 & 300 \\
\hline
Ours & $<60$ & $<120$ & $<180$ \\
\hline
\end{tabular}
\end{center}
\end{table}

\subsection{Vision Model Based Gaussian Splatting}

% Gaussian Splatting is a novel 3D representation method that models the scene using a set of 3D Gaussian points and has a broad range of applications in the field of robotics. Gaussian points are initialized from point cloud. In this work, we use only the image data captured by cameras to initialize the point cloud of the scene.

Gaussian Splatting \cite{3DGS} models the scene using a set of 3D Gaussian points, which are initialized from point cloud. In this work, we use only the image data captured by cameras to initialize the point cloud of the scene.

For a Gaussian point, its attributes include mean $\mu \in \mathbb{R}^3$, opacity $\alpha \in \mathbb{R}$, and covariance $\Sigma$. These attributes can be learned and optimized through backpropagation. The Gaussian function $G(x)$ can be expressed as :

$$
G(x)=(\frac{1}{({2\pi})^{3/2}\det{(\Sigma)}})\exp(-\frac{1}{2}(x-\mu)^T\Sigma^{-1}(x-\mu))
\eqno{(6)}
$$

During image rendering, 3D Gaussians undergo projective transformation into 2D space and the colors are rendered as follows.

$$
C=\sum_{i\in N}c_i\alpha_i' \prod_{j=1}^{i-1}(1-\alpha_j')
\eqno{(7)}
$$
where $c_i$ is the color computed from the spherical harmonic (SH) coefficients of the $i^{th}$ Gaussian, and $\alpha_i'$ is given by evaluating a 2D Gaussian with covariance multiplied by a learned Gaussian opacity. The 2D covariance matrix is calculated by projecting the 3D covariance $\Sigma$ to the camera coordinate system. This differentiable rendering formulation enables an efficient view-dependent appearance modeling through volumetric blending.

Conventional SfMs such as COLMAP exhibit limitations in texture-uniform environments and scenarios with limited view overlap, as discussed in Section I. To address this, we implement MASt3R \cite{MASt3R}, a neural vision model that leverages deep neural networks to jointly estimate camera extrinsic parameter $P$ and reconstruct dense 3D geometry. The framework processes paired images with partial overlap, generating per-pixel depth maps $Z$ alongside confidence map $q$ . Subsequent application of the Weiszfeld algorithm enables a robust estimation of intrinsic camera parameters $K$ using the predicted depth-confidence pairs. Incorporating camera scaling factors $\sigma$, the 3D-to-2D projection operator is formulated as:

$$
Proj(x) = KP\sigma x
\eqno{(8)}
$$

where $x$ is in the point cloud $\chi \in \mathbb{R}^{H \times W \times 3}$. The point observed at pixel $(i,j)$ by $k^{th}$ camera can be calculated as:

$$
\begin{aligned}
\chi_{i,j}^k = x & =Proj^{-1}\left(\sigma_k,K_k,P_k,Z_{i,j}^k\right) \\
& =1/\sigma_k P_k^{-1}K_k^{-1}Z_{i,j}^k[i,j,1]^T
\end{aligned}
\eqno{(9)}
$$

Following point cloud estimation for every image pairs, we implement a global alignment stage that jointly optimizes camera extrinsic parameter $P$ and scale factor $\sigma$ through gradient-based optimization to generate the final point cloud. The scaling factors $\sigma$ and extrinsic matrices $P$ are refined by minimizing multi-view consistency errors:

$$
\left \{ \sigma^*,P^* \right \}=\arg \min_{\sigma,P} \sum_{i,j\in {1,...,n}} q\parallel\chi^i-\chi^j\parallel
\eqno{(10)}
$$
where $q$ denotes the confidence map derived from MASt3R's predictions. After obtaining all $\chi$, we can train GS model initialized with $\chi_{all} = \left\{ \sigma^1 \chi^1 P^1, ..., \sigma^n \chi^n\cdot P^n\right\}$.

\section{Experiments}

To validate the proposed method, we conducted simulations and real-world experiments using a quadrotor drone, demonstrating its effectiveness and applicability on edge devices.

\subsection{Experimental setting}
We set $v_{max}=1 \mathrm{m/s}, yaw_{max}=1 \mathrm{rad/s}$ in Eq.(1) to ensure safe flight, and set $\tau=0.03, thr_{in}=0.04$ in Eq.(5) when calculating FAST features. To address the GPU resource allocation, the simulation environment is deployed on a desktop computer equipped with an NVIDIA RTX 3060 GPU and a 13th Gen Intel(R) Core(TM) i5-13600KF CPU while the physical drone operates using an NVIDIA Jetson Orin NX (16GB) edge processor. Due to the high computational overhead, the comparative algorithms  run on a server with 4 NVIDIA RTX 4090 GPUs and dual Intel Xeon Gold 5318Y CPUs.

\subsection{Simulation Test}

The simulation scenario is a self-built warehouse environment in Gazebo shown in Fig.5(a). The drone starts on the ground and initiates the mapping algorithm after takeoff. The flight trajectory of the drone and the final voxel map are shown in Fig.5(b). 

\begin{figure}[htbp]
\centering
\subfigure[]
{
        \begin{minipage}[b]{.4\linewidth}
        \centering
        \includegraphics[scale=0.116]{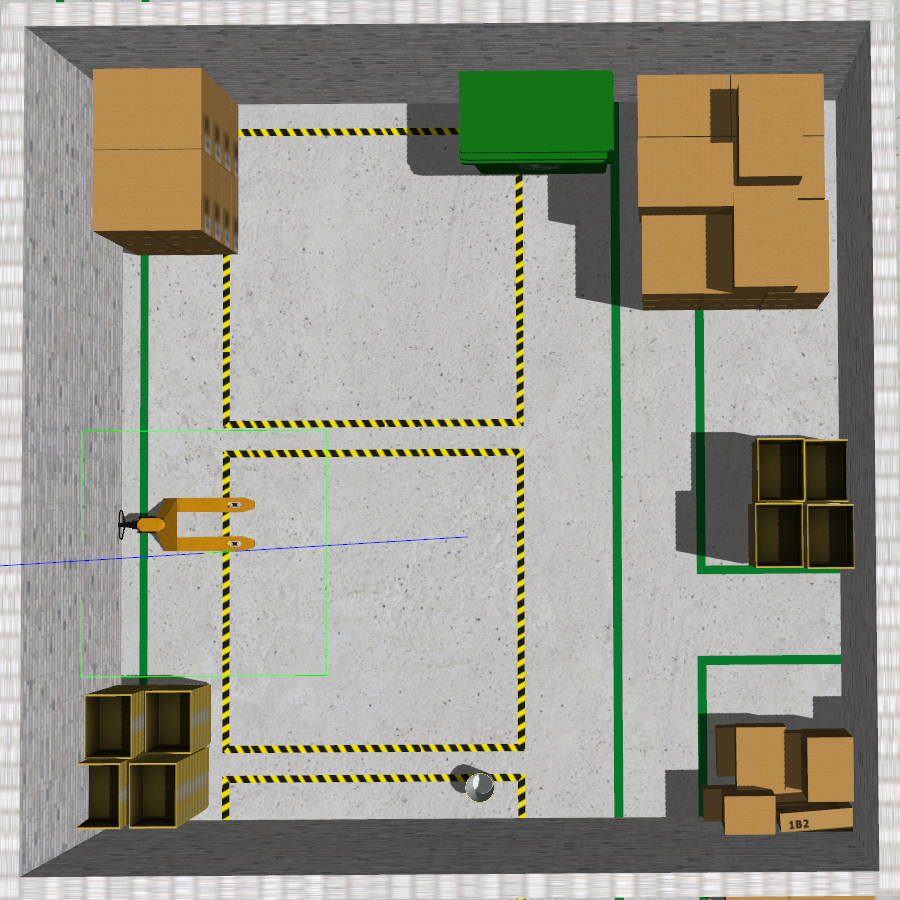}
        \end{minipage}
}
\subfigure[]
{
        \begin{minipage}[b]{.4\linewidth}
        \centering
        \includegraphics[scale=0.158]{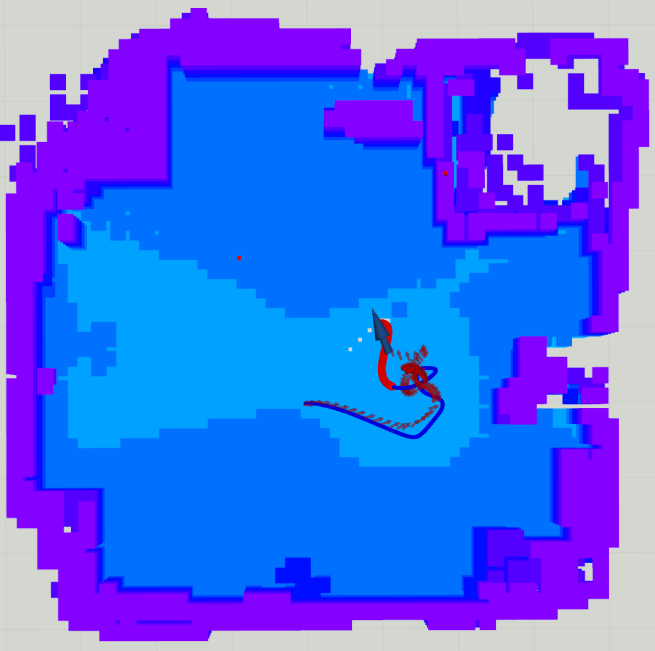}
        \end{minipage}
}
\caption{The gazebo world (a). The process of data collection and exploration in simulation (b).}
\end{figure}

Comparisions in Table II reveal our method is five times faster than the second-fastest approach, while maintaining comparable reconstruction quality. Although InstantSplat achieves the best PSNR result, as shown in Fig.6(b), it reconstructs only a quarter of the scene. The rest of scene is covered due to the wrong initial point cloud, and its PSNR is still high because of the high adaptability of Spherical Harmonics. However, our approach proposed in this paper reconstructs the complete model as shown in Fig.6(a).

\begin{table}[h]
\caption{The reconstruction result in simulation}
\begin{center}
\begin{tabular}{c||c|c}
\hline
Method & PSNR($\uparrow$) & Time($\downarrow$) \\
\hline
% \hline
InstantSplat \cite{InstantSplat} & \textbf{24.12} & 6 minutes \\
% \hline
Splatfactro \cite{Splatfacto} & 21.02 & 14 minutes \\
% \hline
Mip-nerf \cite{Mip-NeRF} & 11.42 & 21 minutes \\
% \hline
Bio-nerf \cite{BioNeRF} & 7.22 & 16 minutes \\
% \hline
Nerf \cite{NeRF} & 16.92 & 25 minutes \\
% \hline
Ours & 23.91 & \textbf{1 minutes} \\
\hline
\end{tabular}
\end{center}
\end{table}

\begin{figure}[htbp]
\centering
\subfigure[]
{
                \begin{minipage}[b]{.4\linewidth}
        \centering
        \includegraphics[scale=0.1]{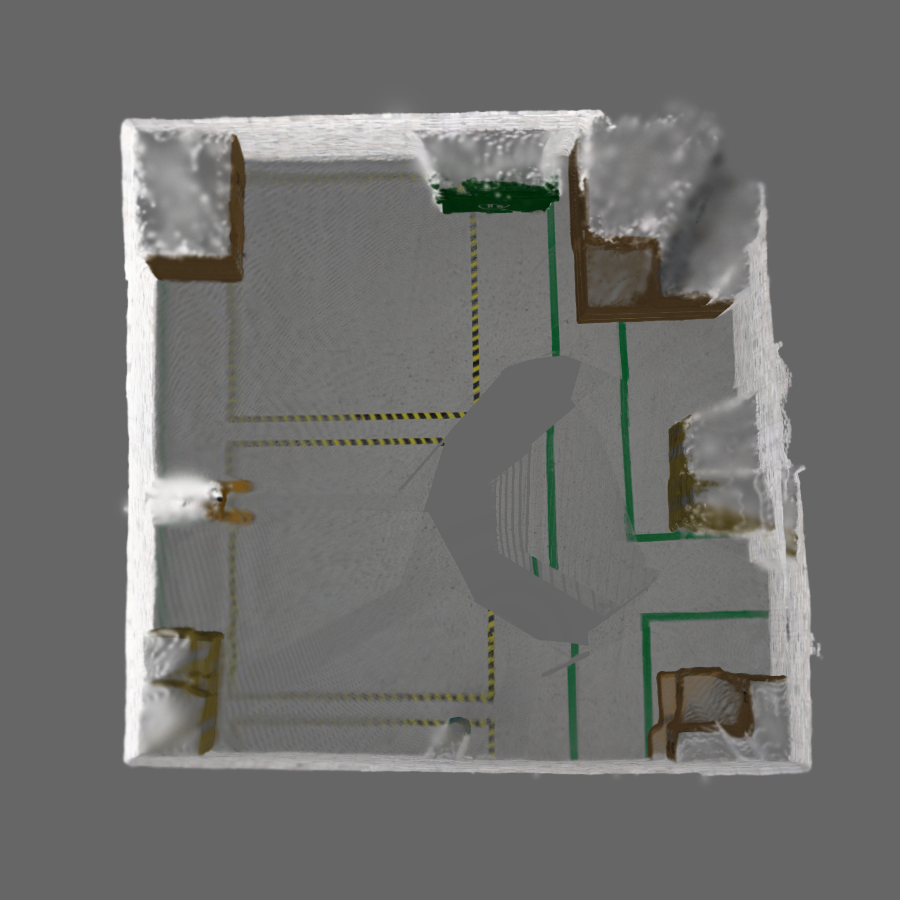}
        \end{minipage}
}
\subfigure[]
{
                \begin{minipage}[b]{.4\linewidth}
        \centering
        \includegraphics[scale=0.1]{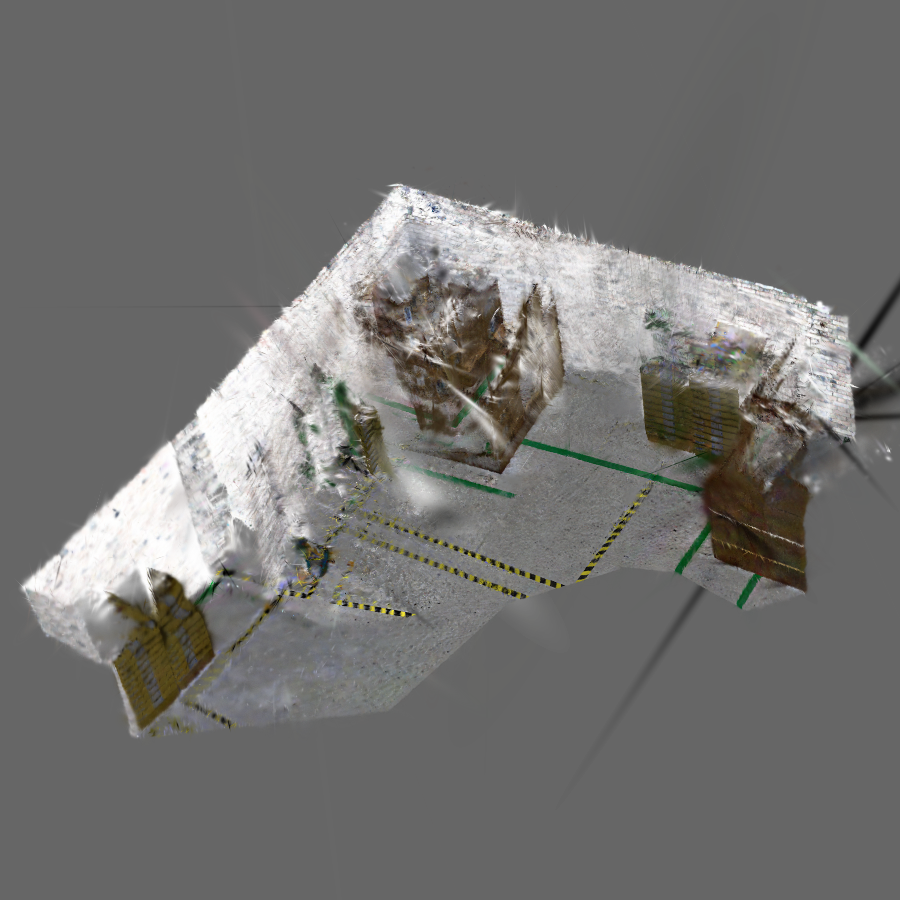}
        \end{minipage}
}
\caption{Gaussian Splatting model reconstructed by Ours (a) and InstantSplat (b). While InstantSplat gets a better PSNR result, it fails to generate the complete model}
\end{figure}

\subsection{Real-World Test}

In the real-world test, as shown in Fig.7, the drone integrates an Intel RealSense D455 depth camera to capture RGB image data. 

\begin{figure}[htbp]
        \centering
        \includegraphics[width=0.6\linewidth]{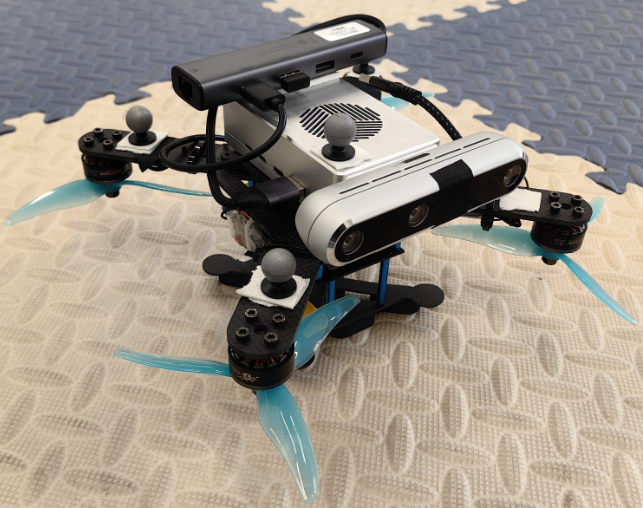}
        \caption{The quadrotor of real-world test. }
\end{figure}

Our experiment is conducted in a room shown in Fig.8(a). We use a PD controller to follow the 
target path, and Fig.8(b) shows the full trajectory during the experiment. The drone performs real-time image pair matching during flight, generating 3D point clouds through visual model inference. Upon completing scene mapping, we conduct a Gaussian Splatting training that produced photorealistic renderings within 1 minute.

\begin{figure}[htbp]
\centering
\subfigure[]
{
        \begin{minipage}[b]{.4\linewidth}
        \centering
        \includegraphics[scale=0.25]{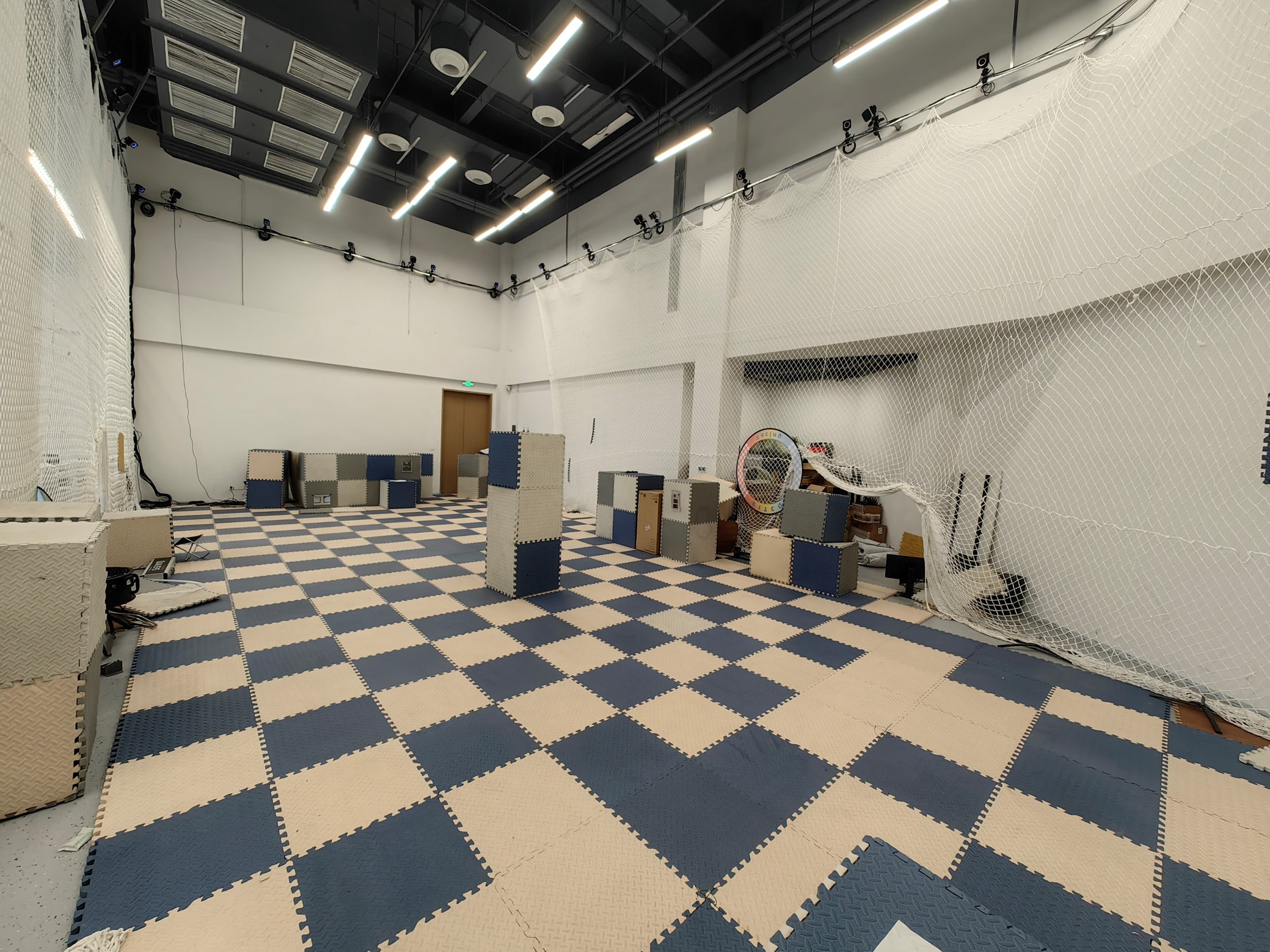}
        \end{minipage}
}
\subfigure[]
{
        \begin{minipage}[b]{.4\linewidth}
        \centering
        \includegraphics[scale=0.25]{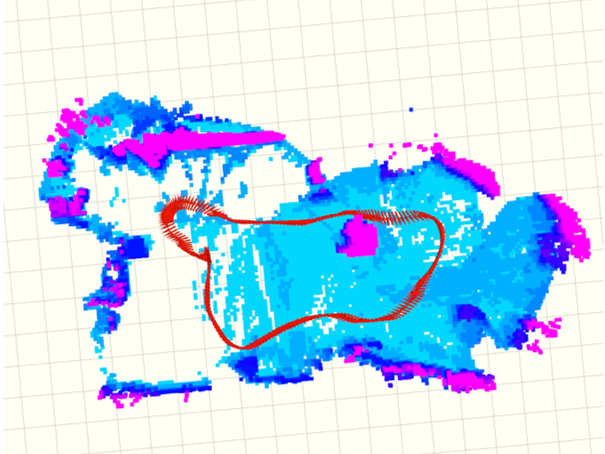}
        \end{minipage}
}
\caption{The real world (a). The process of data collection and exploration in flight (b). The red line is the trajectory of the drone.}
\end{figure}

The experimental results presented in Table III and Fig. 9 demonstrate that both InstantSplat and our proposed method can achieve a complete scene reconstruction. Notably, our approach attains comparable visual quality while requiring merely less than five-sixth of the computational time consumed by InstantSplat.

\begin{table}[h]
\caption{The reconstruction result in real world}
\begin{center}
\begin{tabular}{c||c|c}
\hline
Method & PSNR($\uparrow$) & Time($\downarrow$) \\
\hline
% \hline
InstantSplat & \textbf{25.09} & 6 minutes \\
% \hline
Splatfactro & 23.93 & 13 minutes \\
% \hline
Mip-nerf & 11.42 & 20 minutes \\
% \hline
Bio-nerf & 7.21 & 15 minutes \\
% \hline
Nerf & 8.35 & 25 minutes \\
% \hline
Ours & 24.35 & \textbf{50 seconds} \\
\hline
\end{tabular}
\end{center}
\end{table}

\begin{figure}[htbp]
\centering
\subfigure[]
{
        \begin{minipage}[b]{.4\linewidth}
        \centering
        \includegraphics[scale=0.0955]{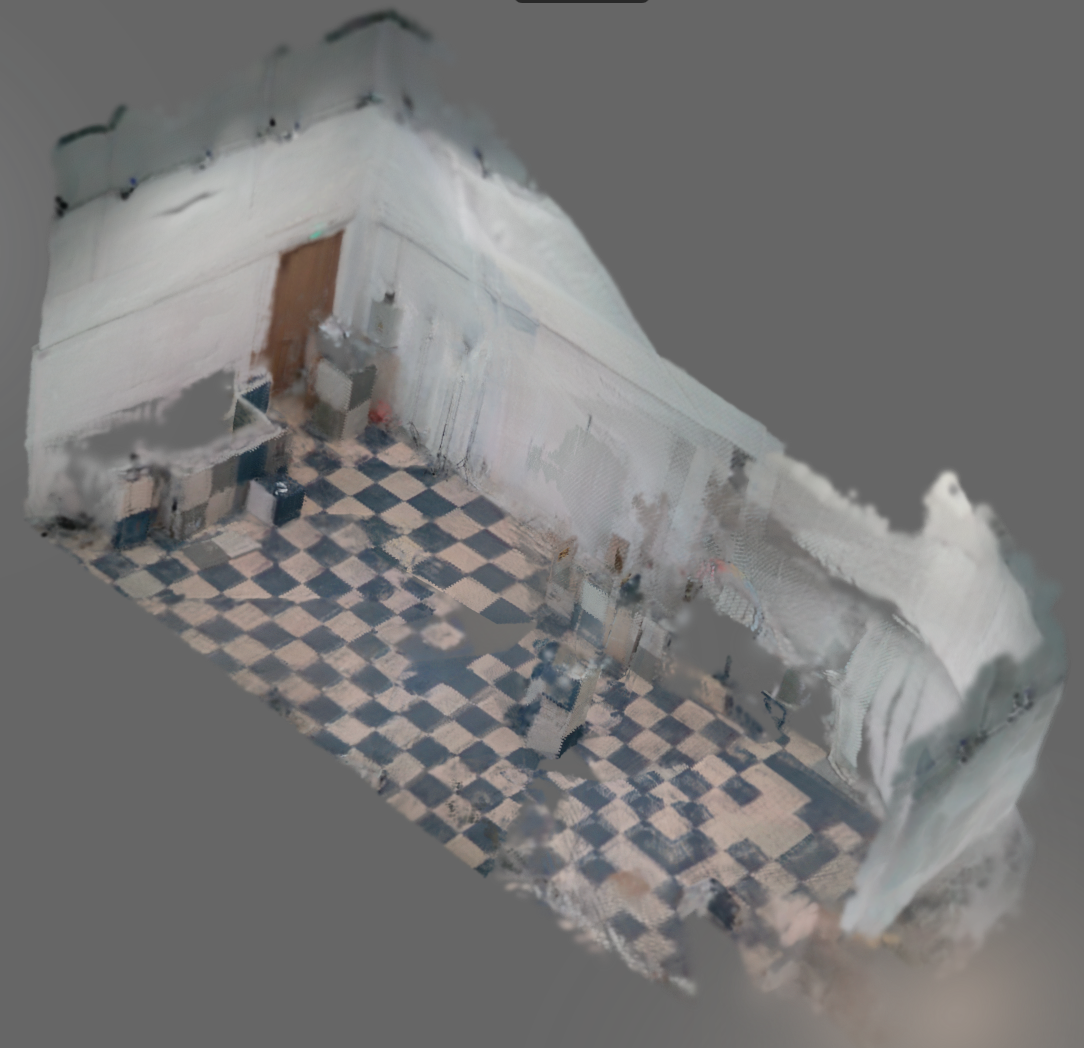}
        \end{minipage}
}
\subfigure[]
{
        \begin{minipage}[b]{.4\linewidth}
        \centering
        \includegraphics[scale=0.09]{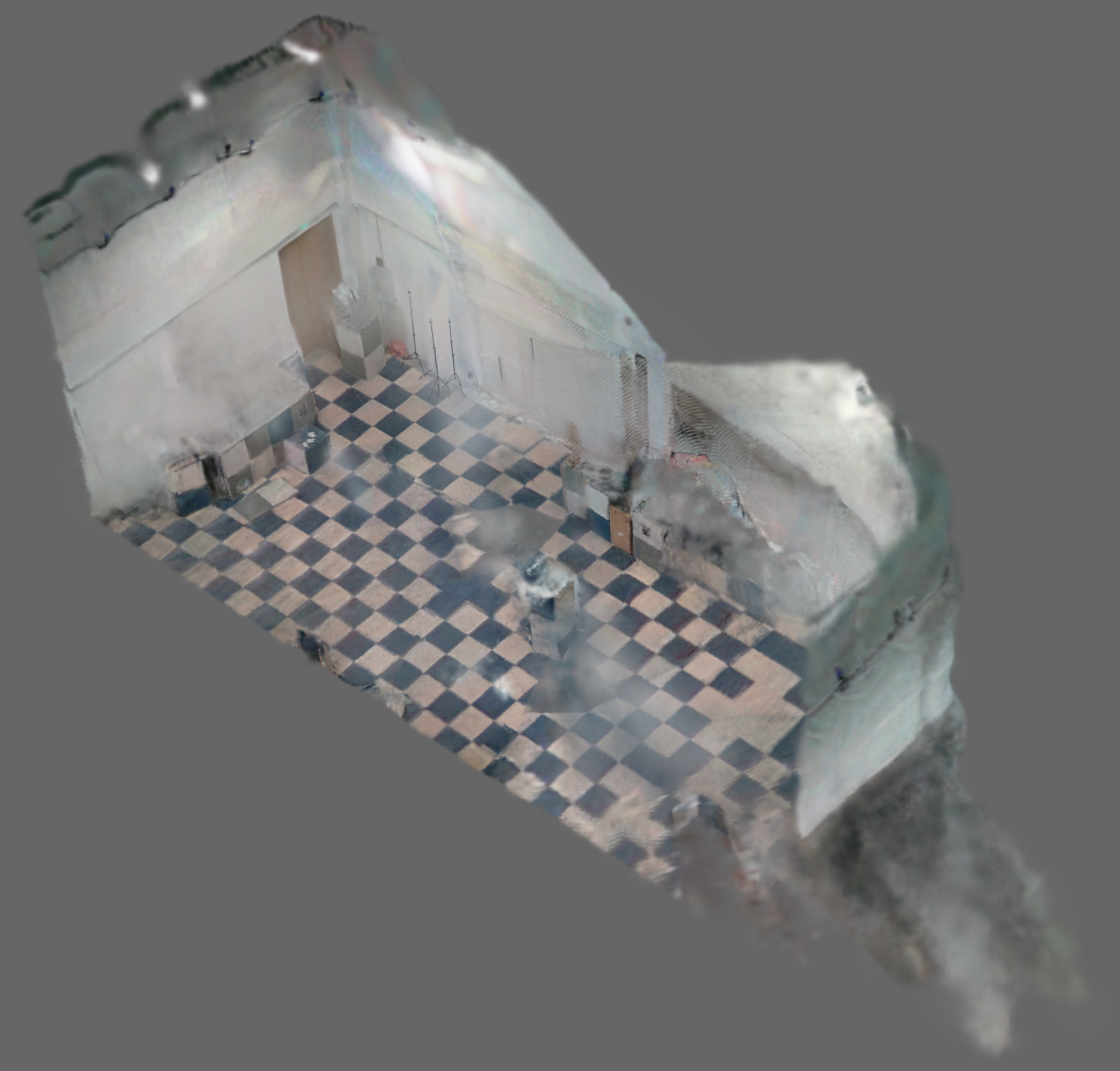}
        \end{minipage}
}
\caption{Gaussian Splatting model rendered by Ours (a) and InstantSplat (b).}
\end{figure}

\section{Conclusion}

This paper has presented a computationally efficient vision-based framework for the 3D scene reconstruction, which achieves a cost-effective environmental perception by replacing the lidar-dependent system with a monocular visual modeling. Extensive validation through simulated scenarios and physical deployments demonstrates its capability to deliver reliable 3D reconstruction performance on resource-constrained embedded platforms, showing better applicability compared to other methods. In future work, instead of a voxel map, we will focus on real-time decision-making and planning in 3DGS representation scenarios.

%%%%%%%%%%%%%%%%%%%%%%%%%%%%%%%%%%%%%%%%%%%%%%%%%%%%%%%%%%%%%%%%%%%%%%%%%%%%%%%%

\bibliographystyle{IEEEtran}

\bibliography{root}

\end{document}